\documentclass{optica-article}

\journal{opticajournal}

\usepackage{graphicx}
\usepackage{braket}
\newcommand{\I}[0]{\mathcal{I}}
\newcommand{\QF}[0]{\mathcal{F}}
\newcommand{\inp}[0]{\mathrm{in}}
\newcommand{\out}[0]{\mathrm{out}}
\newcommand{\eg}[0]{\textit{e.g. }}  

\def\ha{\hat \alpha}
\def\hb{\hat \beta}

\articletype{Research Article}

\usepackage{xcolor}

\begin{document}

\title{Simultaneous quantum estimation of phase and indistinguishability in a two photon interferometer}

\author{Laura T. Knoll,\authormark{1,*} Gustavo M. Bosyk\authormark{2}}

\address{\authormark{1}Laboratorio de Óptica Cuántica, DEILAP, UNIDEF (CITEDEF-CONICET), Buenos Aires, Argentina\\
\authormark{2} Instituto de F\'isica La Plata, CONICET--UNLP, La Plata, Argentina}
\email{\authormark{*}lknoll@citedef.gob.ar} 

\begin{abstract}
With the rapid development of quantum technologies in recent years, the need for high sensitivity measuring techniques has become a key issue. In particular, optical sensors based on quantum states of light have proven to be optimal resources for high precision interferometry. Nevertheless, their performance may be severely affected by the presence of noise or imperfections. In this work we derive the quantum Fisher information matrix associated to the simultaneous estimation of an interferometric phase and the indistinguishability characterizing the probe state consisting of an even number of photons. 
We find the optimal measurement attaining the ultimate precision for both parameters in a single setup and perform an experiment based on a pair of photons with an unknown degree of indistinguishability entering a two-port interferometer. 
\end{abstract}

\section{\label{sec:intro} Introduction}

Understanding what is the ultimate precision limit in estimating a parameter has become an important problem both from a fundamental and industry point of view.
A practical application of quantum mechanics is the measurement of parameters with greater sensitivity than the one achievable in the same experiment performed in a classical setting \cite{giovannetti2011advances, giovannetti2004quantum}.
Using quantum mechanics, advantages over previously known techniques can be obtained for practical tasks such as parameter estimation or state discrimination \cite{toth2014quantum,paris2009quantum}.

In particular, optical metrology uses interferometry as a tool to perform precision measurements \cite{rarity1990two,demkowicz2015quantum,pirandola2018advances,polino2020photonic,barbieri2022optical,berchera2019quantum,genovese2021experimental}. 
The most basic optical interferometer is a two-mode device, whose relative phase difference is unknown. This unknown phase can be designed to encode information about different quantities of interest in different contexts \cite{degen2017quantum,pezze2014quantum,huelga1997improvement,fabre2000quantum,dowling2015quantum,taylor2016quantum}. 
Although photons are an optimal resource in this scenario \cite{giovannetti2011advances,Lang2014,Holland1993,Bollinger1996,Lee2002,aguilar2020robust}, optical sensors based on quantum states of light are susceptible to noise and imperfections that affect their performance.
Therefore, it is necessary to be able to estimate more than one parameter characterizing the dynamics. This problem has important fundamental and technological implications, given that in realistic scenarios losses are inevitable \cite{berchera2019quantum,szczykulska2016multi,vidrighin2014joint,albarelli2020perspective,markiewicz2021simultaneous}.

The problem of phase estimation in noisy environments has already been addressed by taking a multiparameter approach in the context of photon loss/absorption  \cite{crowley2014tradeoff,birchall2020quantum}, phase diffusion \cite{szczykulska2017reaching,vidrighin2014joint} and under limited visibility \cite{roccia2018multiparameter}. 
In a previous work \cite{knoll2019role}, we showed that the quantum Fisher information associated to the phase estimation increases linearly with respect to the degree of indistinguishability between the input photons in a two-port interferometer. This was studied in the ideal case of a pure probe state with a definite even number of photons where the degree of indistinguishability was known.

{Since achieving complete indistinguishability between interfering photons can be a technological challenge, the simultaneous estimation of both phase and indistinguishability allows for a more robust characterization of the problem.
Therefore, in this work we focus on the problem of phase estimation inside an interferometer using a pair of partially indistinguishable photons, where the degree of indistinguishability is also assumed unknown. We follow a multiparameter approach to the problem of estimating simultaneously the phase difference between the two arms of the interferometer, and the degree of indistinguishability which ultimately characterizes the probe state.} 

The process of quantum parameter estimation can be described in three steps:
(i) preparation of an initial quantum state (probe state); 
(ii) parametrization by means of a quantum operation applied to the probe state; 
(iii) measurement of the final state. 
Finally, from the obtained outcome probabilities a classical estimation of the parameter(s) is performed. Although exploiting quantum phenomena can enhance the parameter estimation, the ultimate precision that can be obtained is limited by the quantum Cramér Rao (QCR) bound where the quantum Fisher information (QFI) and quantum Fisher information matrix (QFIM) quantify the precision for single and multi-parameter estimation, respectively.
In the multiparameter estimation problem the QCR bound for a set of parameters $\vec{\text{x}}=\{\text{x}_k \}$ takes the form
\begin{equation}
    \text{Cov}(\vec{\text{x}})\geq \frac{1}{M\mathcal{F_{\vec{\text{x}}}}}
\end{equation}
\label{eq:QCRB}
with $\text{Cov}(\vec{\text{x}})$ the covariance matrix for the set of parameters, $\mathcal{F_{\vec{\text{x}}}}$ the QFIM and $M$ the number of experimental repetitions (see \textit{e.g.} \cite{HelstromBook,liu2019quantum}). The single-parameter case can be retrieved by simply taking the diagonal elements of the QFIM, transforming the matrix inequality \eqref{eq:QCRB} into the scalar QCR bound $(\Delta\text{x}_i)^2\geq (M\mathcal{F}_{ii})^{-1}$ in terms of the QFI $\mathcal{F}_{ii}$. 
For the particular case of a pure parametrized state $|\Psi\rangle:=|\Psi(\vec{\text{x}})\rangle$, the elements of the QFIM can be easily calculated as
\begin{equation}
\label{eq:qfim}
    \mathcal{F}_{ij}=4\textbf{Re} \left( \langle\partial_i\Psi|\partial_j\Psi\rangle-\langle\partial_i\Psi|\Psi\rangle \langle\Psi|\partial_j\Psi\rangle \right),
\end{equation}
where $\ket{\partial_i\Psi}$ is the partial derivative of $\ket{\Psi}$ with respect to $\text{x}_i$.
Multiparameter and single-parameter estimation are fundamentally different.
When dealing with a single parameter, the ultimate bound can always be attained by performing an appropriate measurement. However, in the case of a multi-parameter problem, it may not always be possible to find a measurement that saturates the QCR bound~\eqref{eq:QCRB}.

As mentioned before, we will focus on the problem of simultaneous estimation of phase and indistinguishability, using as a model a probe state consisting of a definite number ($2n$) of partially distinguishable photons entering a two-port interferometer. Essentially, we are interested in characterizing the precision of the estimation, whose ultimate limit is quantified by the QFI. 
Let us first briefly describe the problem of single-parameter phase estimation, assuming complete indistinguishability between the $2n$ photons. For the probe state $\ket{\Psi^\inp} = \ket{2n,0}$, where the $2n$ photons enter the interferometer through one of the ports, the QFI is given by $\mathcal{F}(\ket{\Psi^\inp})=2n$.
This leads to the standard quantum limit (SQL) scaling $\Delta^2(\hat{\phi})= 1/2n$.
This limit can be surpassed by making a better choice of the probe. 
Using the twin-Fock state $\ket{\Psi^\inp} = \ket{n,n}$ as a probe, the QFI is now given by $\mathcal{F}(\ket{\Psi^\inp})=2n(n+1)$ \cite{Lang2014,Holland1993}. For this initial state the phase error scales as the Heisenberg-limit (HL) $\Delta^2(\hat{\phi})\sim 1/2n^2$ (see \eg \cite{Bollinger1996,Lee2002}).
It should be noted that the quantum interference between the $2n$ fully indistinguishable photons entering the interferometer makes it possible to reach this limit.
In real world implementations, this precision can be severely limited by the presence noise, degrading the indistinguishability of interfering photons. In what follows, we will address this problem by taking a multiparameter approach for the simultaneous estimation of both parameters.

\begin{figure}[h!]
\vspace{0.5cm}
  \centering
  \includegraphics[width=0.7\textwidth]{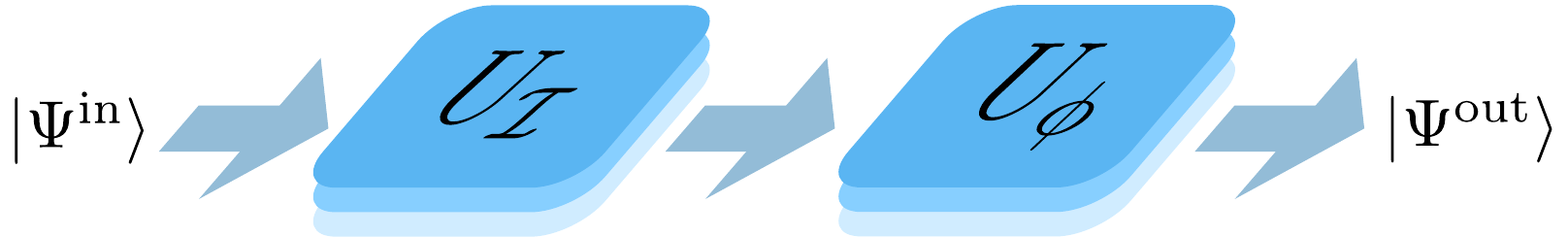}
  \caption{General scheme describing a probe quantum state $\ket{\Psi^{\mathrm{in}}}$ with a definite number of identical photons ($2n$) that loses indistinguishability due to the action of $U_\I$ and then enters a two-mode interferometer given by $U_\phi$ leading to the final state $\ket{\Psi^{\mathrm{out}}}$.}
  \label{fig:scheme}
\end{figure}

Let us describe the general scenario depicted in Fig. \ref{fig:scheme}, where a quantum state $\ket{\Psi^{\mathrm{in}}}$ with a definite number of identical photons (2$n$) loses indistinguishability before entering a two-mode interferometer. 
The degree of indistinguishability $\mathcal{I}$ between interfering photons from different ports, labelled $\alpha$ and $\beta$, is accounted by two extra degrees of freedom, labelled $\mu$ and $\nu$, such that each individual photon is described by a four-mode representation based on annihilation operators $\hat{\alpha}_{\mu}$, $\hat{\alpha}_{\nu}$, $\hat{\beta}_{\mu}$ and $\hat{\beta}_{\nu}$ which satisfy the usual bosonic commutation relations: $[\ha_\mu,\ha^\dag_\nu] = [\hb_\mu,\hb^\dag_\nu] = \delta_{\mu\nu}$ and $[\ha_\mu,\ha_\nu] = [\hb_\mu,\hb_\nu] = 0$.
Let us assume that initially the 2$n$ photons are equally distributed in modes $\hat{\alpha}_{\mu}$ and $\hat{\beta}_{\mu}$.
The unitary map $U_\I$ responsible for encoding the degree of indistinguishability is given by the following transformation
\begin{equation}
\begin{aligned}
\label{eq:Uind}
    & \hat{\alpha}_{\mu} \rightarrow \hat{\alpha}_{\mu}\\
    & \hat{\beta}_\mu \rightarrow \hat{\beta}_\I = \sqrt{\I} \hat{\beta}_\mu + \sqrt{1-\I} \hat{\beta}_\nu\\
    & \hat{\alpha}_{\nu} \rightarrow \hat{\alpha}_{\nu}\\
    & \hat{\beta}_\nu \rightarrow \hat{\beta}_{\I^{-}} = \sqrt{\I} \hat{\beta}_\mu - \sqrt{1-\I} \hat{\beta}_\nu
\end{aligned}
\end{equation}
with $0\leq\I\leq 1$. In this way, the degree of indistinguishability is parametrized by the overlap between photons, where $n$ photons occupy mode $\hat{\alpha}_{\mu}$ while the other $n$ photons are in a superposition of orthogonal modes $\hat{\beta}_{\mu}$ and $\hat{\beta}_{\nu}$.
The phase $\phi$ is then encoded into the $\I$-parametrized state by means of the unitary map $U_{\phi}=\text{exp}(-i\phi H)$
with the Hamiltonian given by
$H= J_\mu \otimes I + I \otimes J_\nu$, where $J_\chi = -i(\hat{\alpha}_\chi^\dag \hat{\beta}_\chi - \hat{\alpha}_\chi \hat{\beta}^\dag_\chi)/2$ with $\chi=\mu,\nu$.
Finally, the fully-parametrized output state can be written as
\begin{equation}
    \label{eq:input}
  \ket{\Psi^{\text{out}}} = U_{\phi} \left( \frac{(\hat{\alpha}_\mu^\dag)^n}{\sqrt{n!}} \frac{(\hat{\beta}_\I^\dag)^n}{\sqrt{n!}} \right) \ket{0}.
\end{equation}

\section{\label{sec:inter} Two-parameter estimation}
\subsection {Quantum Fisher Information matrix for phase and indistinguishability estimation}

We will focus on the two photon case, i.e. $n=1$, such that the initial state is simply $\ket{\Psi^{\text{in}}}=(\hat{\alpha}_\mu^\dag)(\hat{\beta}_\mu^\dag)\ket{0}\equiv\ket{1100}$, where the simplified notation $\ket{{\alpha_\mu} {\beta_\mu}{\alpha_\nu} {\beta_\nu}}$ is used to describe the number of photons in each of the four possible modes.
After encoding the indistinguishability the state becomes $\ket{\Psi_{\I}}=\sqrt{\I}\ket{1100}+\sqrt{1-\I}\ket{1001}$, that is, a superposition between the completely indistinguishable ($\I=1$) and distinguishable ($\I=0$) states.
A phase $\phi$ is then encoded through the action of a phase shift between the arms of a two-port interferometer, usually described by a Mach-Zehnder interferometer consisting on a sequence of a 50:50 beamsplitter followed by a phase shift and another 50:50 beamsplitter.  
Then the final parametrized output state is given by 
\begin{equation}
    \begin{aligned}
    \label{eq:out2phtons}
        \ket{\Psi^\out} & = \sqrt{\I}  \left[\frac{\sqrt{2}\sin(\phi)}{2}(\ket{2000} - \ket{0200}) - \cos(\phi) \ket{1100} \right] \\
        & + \sqrt{1- \I} \left[\frac{\sin(\phi)}{2} \left( \ket{1010} - \ket{0101} \right)- \cos^2(\phi/2)\ket{1001} +\sin^2(\phi/2)\ket{0110}\right].
\end{aligned}
\end{equation}

For this pure parametrized state the calculated QFIM using Eq. \eqref{eq:qfim} results in a diagonal matrix ($\QF_{\I\phi}=\QF_{\phi\I}=0$):
\begin{equation}
\label{eq:qfim_diag}
    \begin{pmatrix}
    \QF_{\phi\phi} & 0\\
    0 & \QF_{\I\I}
    \end{pmatrix}
\end{equation}
with $\QF_{\phi\phi}$ and $\QF_{\I\I}$ the quantum Fisher information for the phase and indistinguishability respectively given by
\begin{equation}
\label{eq:qfi2params}
    \QF_{\phi\phi}=2(\I+1), \quad\quad \QF_{\I\I}=\frac{1}{\I(1-\I)}.
\end{equation}

The QFI corresponding to the estimation of the phase ($\QF_{\phi\phi}$) is independent of the parameter $\phi$ and increases linearly with the degree of indistinguishability, such that for any $\I>0$ one can beat the SQL given by $(\Delta\phi)^2=1/2$. This is in agreement with our previous result for the single-parameter phase estimation problem with sources of photons with a known degree of indistinguishability, as reported in \cite{knoll2019role}.

The QFI for indistinguishability ($\QF_{\I\I}$) is also independent of the phase but does depend on the parameter to be estimated $\I$ through a rational function of degree 2 that is symmetric with respect to $\I$ and diverges for $\I=0,1$. 
These two extreme cases correspond to having either state $\ket{\Psi_{\I}}=\ket{1100}$ for $\I=1$ or state $\ket{\Psi_{\I}}=\ket{1001}$ for $\I=0$, i. e., states with a definite degree of indistinguishability.

To experimentally test these bounds, a measurement $\mathcal{M} = \{M_m\}$ has to to be performed to obtain the outcome probabilities $p(m|\I,\phi) = \braket{\Psi^\out|M^\dag_m M_m|\Psi^\out}$ from which the classical Fisher information matrix (FIM) can be calculated:
\begin{equation}
    F_{ij}=\sum_{m \in \mathfrak{M}} \frac{ \left[\partial_i p(m|\vec{\text{x}})\right] \left[\partial_j p(m|\vec{\text{x}})\right] }{p(m|\vec{\text{x}})}.
\end{equation}

It is clear that the FIM depends on the choice of measurement as it is obtained from the measured probability distribution. In general, a measurement that can attain the QFIM may not exist \cite{liu2019quantum}, unlike the case of single-parameter estimation. 
However, for the problem of estimating both the phase and indistinguishability in a two-photon interferometer, we found that by performing the projective measurements $\mathcal{M}=\{\ket{m}\bra{m}\}_{m \in \mathfrak{M}}$, with 
$\mathfrak{M} = \{2000,0200,1100,1010,\allowbreak 0101,1001,0101, 0011,0020,0002\}$ the corresponding Fisher information matrix is equal to the QFIM given by Eqs. \eqref{eq:qfim_diag} and \eqref{eq:qfi2params}. More precisely, the outcome probabilities $p(m|\I,\phi)$ are given by:
\begin{equation}
\begin{aligned}
\label{eq:probas}
& p(2000|{\I},\phi)= p(0200|{\I},\phi) = \I \frac{\sin^2(\phi)}{2}, \\
& p(1100|{\I},\phi) = \I \cos^2(\phi), \\
& p(1010|{\I},\phi)= p(0101|{\I},\phi) = (1-\I) \frac{\sin^2(\phi)}{4}, \\
& p(1001|{\I},\phi)= (1-\I) \cos^4(\phi/2), \\
& p(0110|{\I},\phi) = (1-\I) \sin^4(\phi/2), \\
& p(0020|{\I},\phi)= p(0002|{\I},\phi) = p(0011|\phi) =0,
\end{aligned}
\end{equation}
from which we obtain
\begin{equation}
\label{eq:fi2params}
\begin{aligned}
  & F_{\phi\phi} =  \sum_{m \in \mathfrak{M}} \frac{\left|\partial_{\phi}p(m|\I,\phi)\right|^2}{p(m|\I,\phi)}= 2(\I+1) = \QF_{\phi\phi},\\
  & F_{\I\I}=\sum_{m \in \mathfrak{M}} \frac{\left|\partial_{\I}p(m|\I,\phi)\right|^2}{p(m|\I,\phi)}= \frac{1}{\I(1-\I)} = \QF_{\I\I},\\
  & F_{\I\phi}= 0 =\QF_{\I\phi}, \\
  & F_{\phi\I}= 0=\QF_{\phi\I}.
\end{aligned}
\end{equation}

Interestingly, this projective measurement corresponds to the optimal measurement providing the ultimate precision for the simultaneous estimation of both parameters, for the given probe state and parameter encoding. This is a notable result, demonstrating that a unique optimal measurement is capable of attaining the quantum limits for the two parameters simultaneously, making it possible to estimate both the phase and indistinguishability with a single probe state. 

\subsection{Two-photon interferometer}

\begin{figure}[h!]
\vspace{0.5cm}
  \centering
  \includegraphics[width=0.85\textwidth]{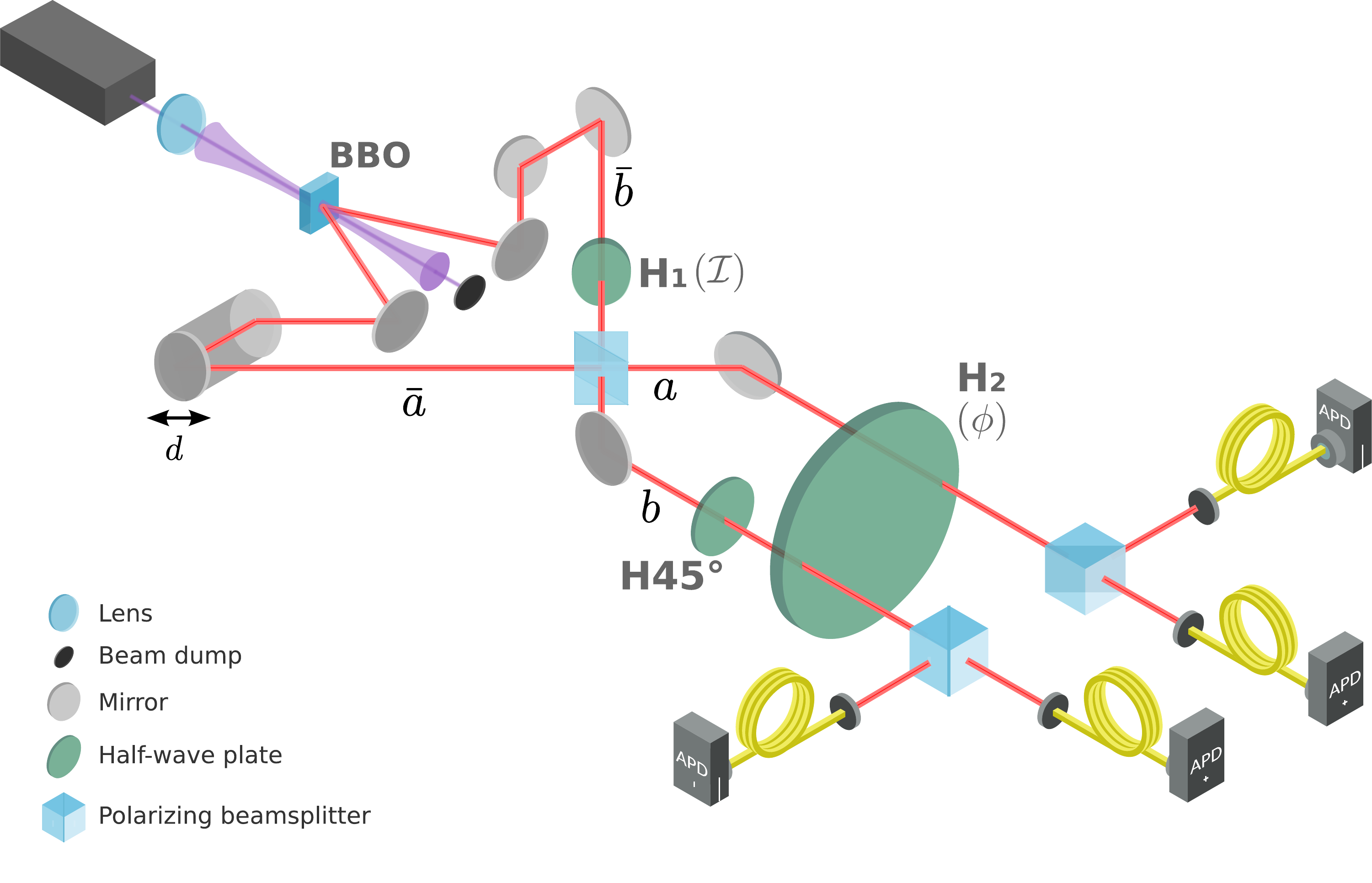}
  \caption{Experimental setup: pairs of horizontally polarized photons at 810nm are generated by Type-I SPDC and directed into path modes $\bar{a}$ and $\bar{b}$. A half-wave plate (HWP) H1 in path $\bar{b}$ encodes the degree of indistinguishability, by making both photons exit the polarizing beamsplitter (PBS) onto the same path mode $a$ or onto orthogonal modes $a$ and $b$. The photons after the PBS have orthogonal polarizations $h$ and $v$ (this is achieved on path mode $b$ by the use of a HWP at $45^{\circ}$). HWP H2 acts as a polarization interferometer, adding a relative phase between the two polarization modes. Measurements are performed by recording coincidences between the four possible output modes $h_a,v_a,h_b,v_b$. Each output mode is coupled into a single mode optical fiber and connected to single photon detectors.}
  \label{fig:setup}
\end{figure}

We perform an experiment to study the QCR bound in terms of the QFIM \eqref{eq:qfim_diag} for both parameters $\phi$ and $\I$ simultaneously. The experimental setup is depicted in Figure \ref{fig:setup}, where pairs of photons enter a polarization-based interferometer, such that orthogonal polarization modes $h$ (horizontal) and $v$ (vertical) play the role of interfering modes $\alpha$ and $\beta$ described previously, while the indistinguishability is encoded on path modes $a$ and $b$. 

The photon pairs are generated by spontaneous parametric down-conversion (SPDC) on a type-I BBO non-linear crystal. By pumping the crystal with a 405nm vertically polarized CW laser, signal and idler photons are generated at 810nm with horizontal polarization and directed into path modes $\bar{a}$ and $\bar{b}$. Both photons are then recombined in a polarizing beamsplitter (PBS). The distance travelled by photons in path $\bar{a}$ is adjusted with a translation stage so that both photons arrive at the PBS at the same time, guaranteeing indistinguishability. Photons travelling in path mode $\bar{b}$, before passing through the PBS, go through a half-wave plate (HWP), labelled as H1 in Fig. \ref{fig:setup}, that rotates their polarization into a superposition of horizontal and vertical polarization such that $h\mapsto \cos(2\varphi)h+\sin(2\varphi)v$. Then, horizontally polarized photons in path $\bar{b}$ will go through the PBS transmitted into path mode $b$ whereas photons with vertical polarization in path $\bar{b}$ will be reflected into path mode $a$. On the other hand, horizontally polarized photons in path mode $\bar{a}$ are left unchanged and transmitted by the PBS into path mode $a$. The degree of indistinguishability {$\I$} is controlled by adjusting the angle $\varphi$ of HWP H1, making both photons exit the PBS onto the same path mode $a$ (complete indistinguishability) or onto orthogonal modes $a$ and $b$ (complete distinguishability). 
An additional HWP at $45^{\circ}$ is placed on path $b$ to rotate the horizontally polarized photons into vertical ones such that the final transformation can be written as
\begin{equation}
    \begin{aligned}
    & h_{\bar{a}}\mapsto h_a\\
    & h_{\bar{b}}\mapsto \sqrt{\I} v_a + \sqrt{1-\I} v_b
    \end{aligned}
\end{equation}
which is equivalent to \eqref{eq:Uind} with $\I=\sin^2(2\varphi)$. In this way, we obtain the desired two-photon $\I$-parametrized state $\ket{\Psi_\I}=\sqrt{\I}\ket{1_h1_v00}+\sqrt{1-\I}\ket{1_h001_v}$.

The phase $\phi$ is then encoded onto this state by the action of another half-wave plate (H2 in Fig.\ref{fig:setup}), acting on both path modes $a$ and $b$ as a polarization interferometer. This HWP performs the transformation $h_a \mapsto \cos(2\theta) h_a+\sin(2\theta) v_a$ and $v_i \mapsto \sin(2\theta) h_i - \cos(2\theta) v_i$ with $i=a,b$ and where $\theta=\phi/4$ is the physical angle of the HWP.
Therefore, the state at the output of the interferometer takes the same form as the one on Eq.\eqref{eq:out2phtons}.

Measurements of the outcome probabilities $\{p(m|{\I},\phi) = |\braket{m|\Psi^\out}|^2\}_{m \in \mathfrak{M}}$ are performed through the projective measurement
$\mathcal{M}=\{\ket{m}\bra{m}\}_{m \in \mathfrak{M}}$, with $\mathfrak{M} = \{2_h000,02_v00,1_h1_v00,\allowbreak 1_h01_h0, 01_v01_v,1_h001_v,01_v01_v, 001_h1_v,002_h0,0002_v\}$ (although projections onto the last three elements do not contribute, see \eqref{eq:probas}).
This measurements are performed by placing PBSs at the end of each path mode $a$ and $b$, projecting onto polarization modes $h$ and $v$, and using four single photon detectors, one for each possible outcome $h_a,v_a,h_b,v_b$. Coincidences are measured between the corresponding detectors according to each probability measurement. Each output mode is coupled into a single-mode fiber and 10nm interference filters are placed before the optics corresponding to exit modes $h_a,v_a$ while band-pass filters are used for outputs $h_b,v_b$.
Projective measurements onto $\{2_h000,02_v00\}$ are obtained by including 50:50 fiber beamsplitters (FBS) at the output modes $h_a$ and $v_a$ and measuring coincidences between the outputs of each FBS.

\section {Experimental results}

As mentioned in the previous section, initial indistinguishability is guaranteed between the two photons generated by SPDC by adjusting the optical path length of mode $\bar{a}$ so that both photons overlap at the PBS. A polarization Hong-Ou-Mandel interferometer \cite{hong1987,Birchall2016,Slussarenko2017,Matthews2016,knoll2019role} is obtained for the setting $\varphi=\frac{\pi}{4}$ and $\phi=\frac{\pi}{2}$ ($\theta=\frac{\pi}{8}$).
Then, by measuring the coincidences between detectors $h_a$ and $v_a$ as a function of the path length difference we observe the two-photon interference effect and identify the path length which guarantees maximum indistinguishability, achieving an interference visibility of $\mathcal{V}=(87\pm 1)\%$.

The experimental data were obtained from the projective measurements  $\mathcal{M}$ for $45$ equally spaced  values of the phase $\phi$ in the range $[0,\pi]$ and for $5$ values of indistinguishability in the range $[0,1]$.
For each indistinguishability value we obtain the normalized coincidence count rates $n_{m_i}/N$ for the possible outcomes $m_i \in \{m_1=2_h000,m_2=02_v00,m_3=1_h1_v00, m_4=1_h01_h0, m_5=01_v01_v,m_6=1_h001_v,\allowbreak m_7=01_v01_v\}$, with $N=750$ the mean number of total photons detected (projections onto $\{002_h0,0002_v,001_h1_v\}$ were not performed, since the corresponding theoretical probabilities are null).

\begin{figure}[h!]
  \centering
  \includegraphics[width=\textwidth]{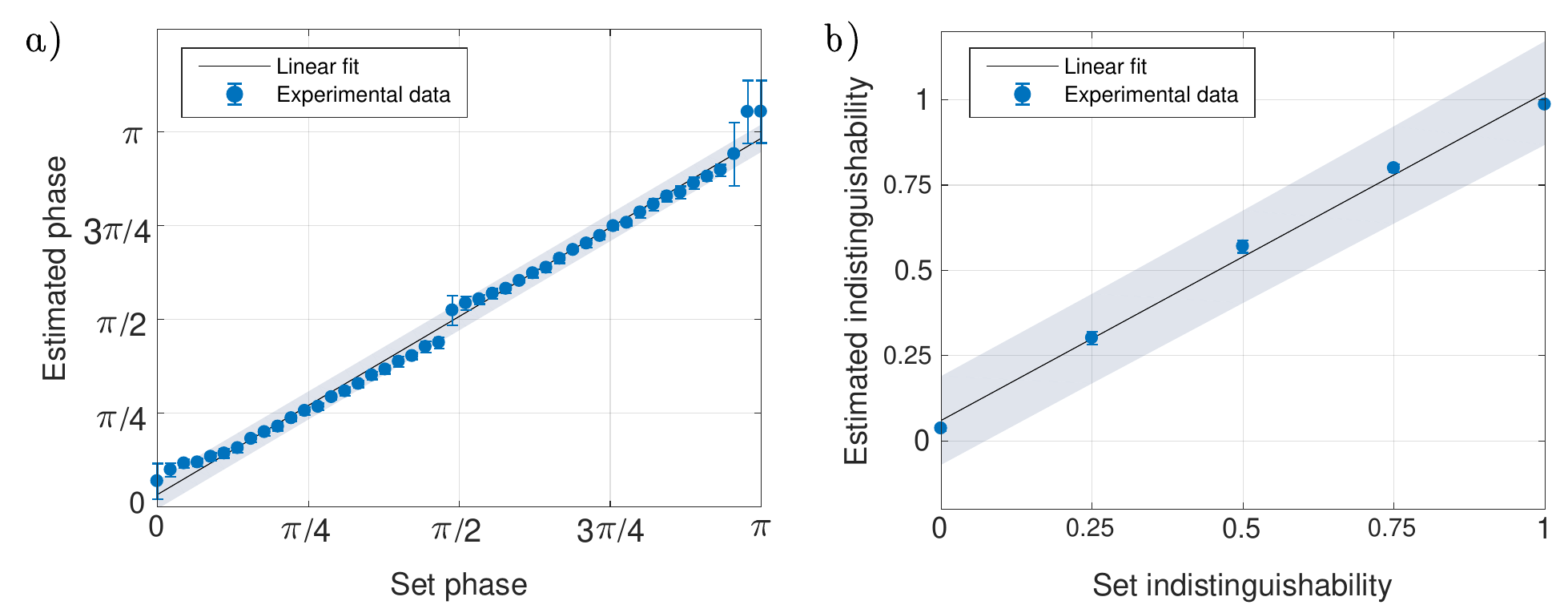}
  \caption{Estimated a) phase and b) indistinguishability as a function of the corresponding set parameters. Symbols represent the experimental data. The solid black line corresponds to a least square linear fit function $y=a_1x+a_2$ with parameters a) $a_1=0.95\pm 0.03$ and $a_2=0.026\pm0.012$; and b) $a_1=0.96\pm 0.13$ and $a_2=0.06\pm0.08$.  
  The shaded area corresponds to the $68\%$ confidence limits for the least square fit.}
  \label{fig:estimates}
\end{figure}

For each pair $(\I,\phi)$ we simulate $M=10^4$ experiments sampling $N=750$ times, where the simulated count rates $x_{m_i}$ are distributed according to the multinomial distribution 
\begin{equation}
   P(x_{m_1},\cdots,x_{m_7}) = \frac{N!}{x_{m_1}! \cdots x_{m_7}!} \prod_{i=1}^7 p_{m_i}^{x_{m_i}}
\end{equation}
with $p_{m_i} = n_{m_i}/N$, $\sum_i p_{m_i} = 1$ and $\sum_i x_{m_i} = N$.
For each simulated experiment, we obtain the maximum-likelihood (ML) estimates $(\hat{\phi}_\ell,\hat{\I}_\ell)$ with $\ell=1, \ldots, M$, assuming that the simulated counts follow a multinomial distribution with the theoretical probabilities as expressed in Eq. \eqref{eq:probas}. 
Averaging over all $M$, we obtain the estimated values $(\hat{\phi},\hat{\I})$ for the indistinguishability and phase with their corresponding variances $(\Delta^2\hat{\phi},\Delta^2\hat{\I})$. 
Figure \ref{fig:estimates} shows the estimated values for a) the phase and b) the indistinguishability as a function of the respective set parameters. In both plots, the symbols represent the experimental data while the solid black line corresponds to a least square linear fit with its corresponding $68\%$ confidence limits represented in the shaded grey area.
The linear fit function $y=a_1x+a_2$ parameters are given by a) $a_1=0.95\pm 0.03$ and $a_2=0.026\pm0.012$; and b) $a_1=0.96\pm 0.13$ and $a_2=0.06\pm0.08$. 

Finally, averaging the corresponding variances over all values of $\phi$, we obtain the quantities $\braket{F^{ML}_{\phi}}=1/\big(N \braket{\Delta^2\hat{\phi}} \big)$ and $\braket{F^{ML}_{\I}}=1/\big(N\braket{\Delta^2\hat{\I}} \big)$ for each value of the set indistinguishability $\I$. 
Therefore, $\braket{F^{ML}_{\phi}}$ and $\braket{F^{ML}_{\I}}$ are the Fisher information computed from ML estimates, since the ML estimators are unbiased and saturate the Cramér-Rao bound in the asymptotic limit, similar to that discussed in~\cite{Matthews2016}.
These quantities can be also compared to the diagonal elements of the QFIM Eq. \eqref{eq:qfi2params}.

\begin{figure}[h!]
  \centering
  \includegraphics[width=\textwidth]{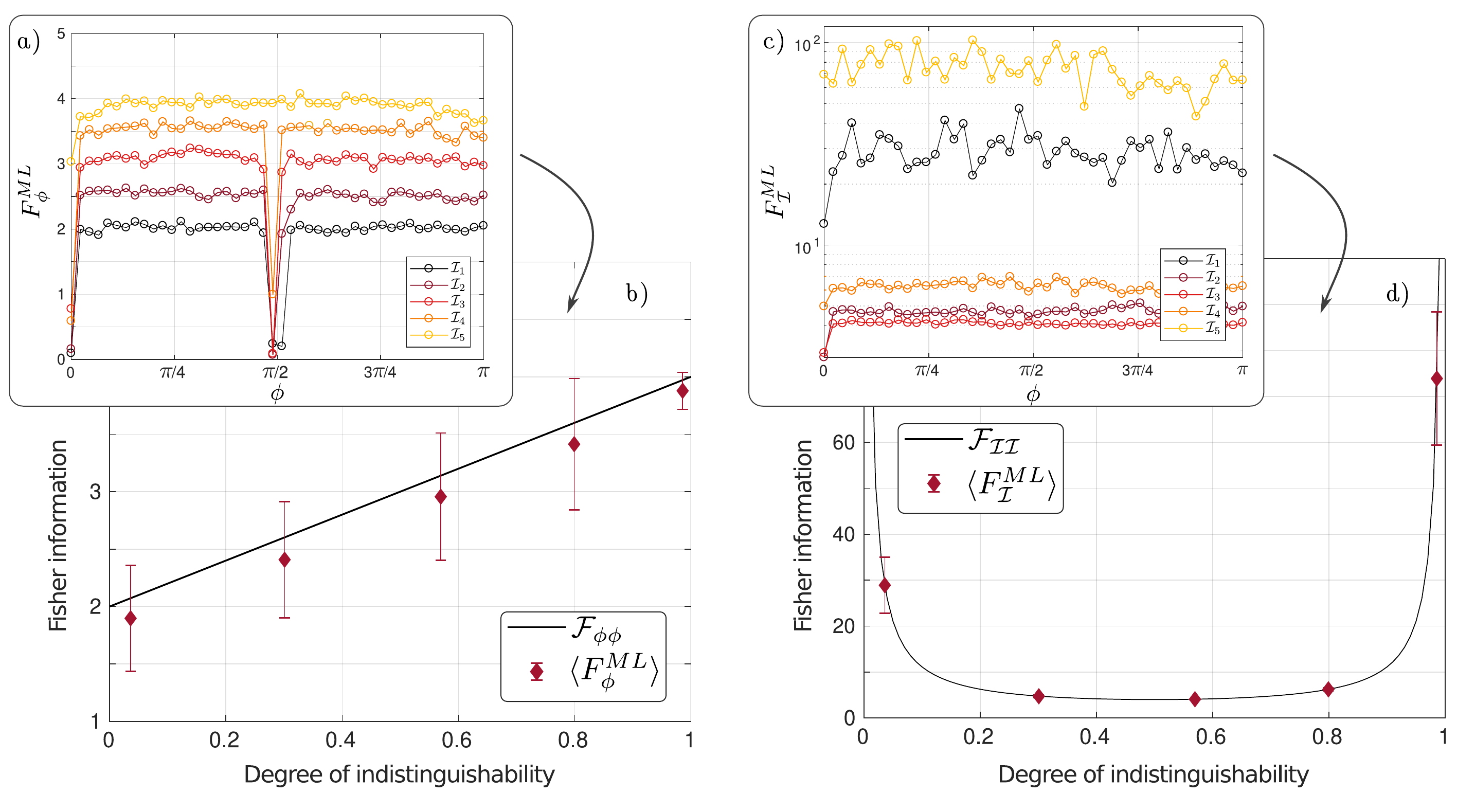}
  \caption{Experimental results. a) Fisher information ${F^{ML}_{\phi}}$ obtained from the maximum-likelihood estimation as a function of the phase $\phi$ for five different values of indistinguishability, in increasing order (black to yellow); b) symbols represent the averaged $\braket{F^{ML}_{\phi}}$ over all phase values for each estimated indistinguishability value while the solid line corresponds to the QFI $\QF_{\phi\phi}$ as a function of the degree of indistinguishability. Analogously c) shows the Fisher information ${F^{ML}_{\I}}$ as a function of the phase $\phi$ for five different values of indistinguishability, in increasing order (black to yellow) and in d) symbols represent the averaged $\braket{F^{ML}_{\I}}$ over all phase values for each estimated indistinguishability value while the solid line corresponds to the QFI $\QF_{\I\I}$ as a function of the degree of indistinguishability.}
  \label{fig:results}
\end{figure}

Figure \ref{fig:results}a) shows the value of $F^{ML}_{\phi}$ as a function of the phase $\phi$ for each set value of $\I$. This value is independent of the phase (aside from the dips observed at $\phi=0$ and $\phi=\pi/2$, where some of the probabilities \eqref{eq:probas} are equal to zero, giving place to biased estimators) and increases with the value of indistinguishability. Averaging over all phase values we can see the behaviour of ${\braket{F^{ML}_{\phi}}}$ as a function of the degree of indistinguishability, as shown in Fig. \ref{fig:results}b). Symbols correspond to $\braket{F^{ML}_{\phi}}$ for each estimated $\hat{\I}$, while the solid line represents the value of $\mathcal{F}_{\phi\phi}$ given by Eq. \eqref{eq:qfi2params} showing a good agreement between theory and experiment. 
In a similar manner, Fig. \ref{fig:results}c) shows the value of $F^{ML}_{\I}$ as a function of the phase $\phi$ for each set value of $\I$, which is again independent of the phase. Averaging over all phase values we obtain the behaviour of ${\braket{F^{ML}_{\I}}}$ as a function of the degree of indistinguishability, as shown in Fig. \ref{fig:results}d). Symbols represent the value of $\braket{F^{ML}_{\I}}$ for each estimated $\hat{\I}$, while the solid line corresponds the value of $\mathcal{F}_{\I\I}$ given by Eq. \eqref{eq:qfi2params} showing also a good agreement between theory and experiment.

\section{\label{sec:final} Final remarks and discussion}

By taking a multiparameter approach we have studied the problem of simultaneously estimating the indistinguishability between pairs of photons and the phase difference between them inside a two-port interferometer. We derived the quantum Fisher information matrix 
which characterizes the ultimate precision attainable. We also found the optimal projective measurement that saturates the Cramér-Rao bound for both parameters simultaneously. To test this bound, we performed an experiment where two initially identical photons go through a transformation that degrades their degree of indistinguishability before entering a two-port interferometer where an unknown phase is encoded. Our theoretical and experimental results contribute to the relevant problem of multiparameter estimation, characterizing simultaneously parameters related to the sensing object and noise affecting the probe state in a single setup.

\begin{backmatter}
\bmsection{Funding}

\bmsection{Acknowledgments}
LTK would like to thank Miguel A. Larotonda and Agustina G. Magnoni for stimulating and helpful discussions.  

\bmsection{Disclosures}
The authors declare no conflicts of interest.

\bmsection{Data availability} Data underlying the results presented on this paper are not publicly available at this time but may be obtained from the authors upon reasonable request.

\end{backmatter}



\providecommand{\noopsort}[1]{}\providecommand{\singleletter}[1]{#1}%

\end{document}